\documentclass[sigconf,nonacm]{acmart}
\AtBeginDocument{%
  }

\usepackage{enumitem}
\usepackage{pifont}
\usepackage{makecell}
\usepackage[framemethod=TikZ]{mdframed}
\usepackage{multirow}
\usepackage{anyfontsize}
\usepackage{adjustbox}
\usepackage[labelfont=bf,textfont=md]{caption}
\usepackage{rotating}
\usepackage{hhline}
\usepackage{threeparttable}
\usepackage{ifthen}
\usepackage{xspace}
\usepackage{booktabs}
\usepackage{colortbl}
\usepackage[dvipsnames]{xcolor}
\usepackage{balance}

\usepackage[most]{tcolorbox}

\definecolor{mygray}{RGB}{247,247,247}

\newtcolorbox[auto counter]{promptbox}[2][]{%
  enhanced,
  colback=mygray,
  colframe=black,
  fonttitle=\bfseries,
  title={Prompt \thetcbcounter: #2},
  boxrule=1pt,
  arc=4pt,
  left=2mm, right=2mm,
  top=1mm, bottom=1mm,
  #1
}

\newlist{rqlist}{enumerate}{1}
\setlist[rqlist]{label*=RQ\arabic*~, ref=RQ\arabic*}

\newcommand{\cmark}{\ding{51}}
\newcommand{\xmark}{\ding{55}}

\newcommand{\rqanswerbox}[1]{%
  \begin{tcolorbox}[colback=mygray,colframe=black,lowerbox=invisible,savelowerto=\jobname_ex.tex]
    #1
  \end{tcolorbox}
}
\mdfsetup{skipabove=\topskip,skipbelow=\topskip}

\newcommand{\link}[1]{\href{#1}{\textit{\textcolor{purple}{#1}}}}
\newcommand{\repourl}{\link{https://github.com/itsluketwist/summary-mediated-repair/}}

\newcommand{\hepack}{\texttt{HumanEvalPack}\xspace}
\newcommand{\humaneval}{\texttt{HumanEval}\xspace}
\newcommand{\mbpp}{\texttt{MBPP}\xspace}

\newcommand{\function}[1]{\textsf{\textsc{#1}}\xspace}
\newcommand{\repair}{\function{Repair}}
\newcommand{\test}{\function{Test}}
\newcommand{\summarise}{\function{Summarise}}
\newcommand{\generate}{\function{Generate}}

\renewenvironment{quote}{%
  \list{}{%
    \leftmargin0.5cm   
    \rightmargin\leftmargin
  }
  \item\relax
}
{\endlist}

\newcommand{\prompt}[1]{
    \begin{quote}
    \textit{
        ``#1''
    }
    \end{quote}
}

\begin{document}

\title[Code Summaries as Diagnostic Context for LLM-Based Program Repair]{Code Summaries as Diagnostic Context\\for LLM-Based Program Repair}

\author{Lukas Twist}
\orcid{0009-0009-6640-2532}
\affiliation{%
  \institution{King's College London}
  \city{London}
  \country{UK}
}
\email{lukas.twist@kcl.ac.uk}

\author{Jie M. Zhang}
\orcid{0000-0003-0481-7264}
\affiliation{%
  \institution{King's College London}
  \city{London}
  \country{UK}
}
\email{jie.zhang@kcl.ac.uk}

\renewcommand{\shortauthors}{Twist and Zhang}

\begin{abstract}

LLMs can generate useful code, but their outputs often contain small implementation-level bugs with large behavioural effects.
In this paper, we ask whether natural-language code summaries provide useful diagnostic context for repairing these errors.
We use \textit{summary-mediated repair} as a simple prompt-only test of their diagnostic value: an LLM first summarises buggy code, then generates a candidate repair conditioned on that summary.
We evaluate \textit{summary-mediated repair} across eight LLMs in two function-level repair settings: existing bugs from \hepack and LLMs' own failed generations of \mbpp.
Diagnostic, error-aware summaries perform best, repairing up to 65\% of previously unseen errors and improving over direct repair by 5\% on average.
However, overall gains are modest and model-dependent, and summaries do little to overcome the difficulty of prompt-only self-repair.
Overall, our results suggest that code summaries are a useful lightweight diagnostic layer for LLM-based program repair, but not a complete repair method on their own.

\end{abstract}

\maketitle


\begin{figure*}
    \centering
    \includegraphics[width=\textwidth]{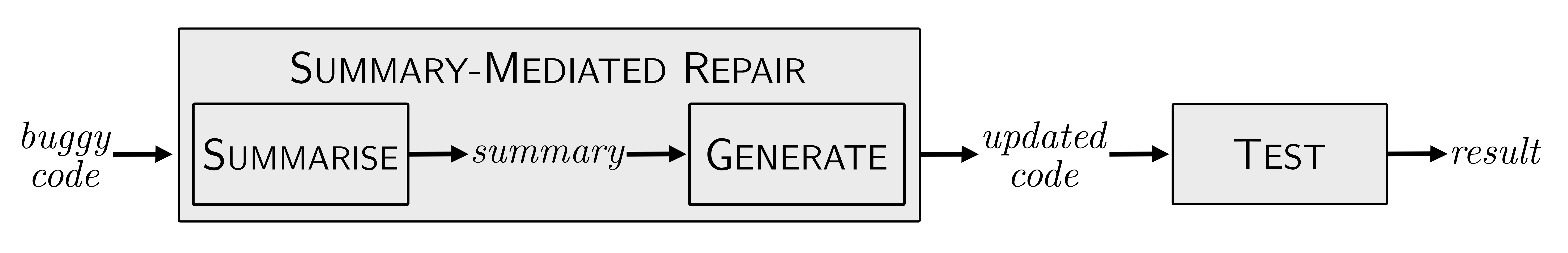}
    \caption{
        \textbf{\textit{Summary-Mediated Repair.}}
        The LLM first summarises the buggy code, generating a natural-language intermediate artefact.
        It then generates a repaired implementation conditioned on that summary, before the candidate repair is validated using tests.
    }
    \Description{The flow...}
    \label{fig:method}
\end{figure*}

\section{Introduction}\label{sec:intro}

Large Language Models (LLMs) have made rapid progress on code generation and benchmark performance~\cite{chenSurveyEvaluatingLarge2024,jiangSurveyLargeLanguage2024}, but they still commonly produce subtle implementation-level bugs, including off-by-one errors, incorrect boundary checks, and small interface misuses~\cite{liuYourCodeGenerated2023}.
These failures are easy to miss, but can substantially change program behaviour, making them an important target for automated program repair (APR)~\cite{yangSurveyLLMbasedAutomated2025}.
For example, using $<$ instead of $<=$, or missing $+1$ in a loop condition, may be enough to make an otherwise plausible function fail.

These errors are increasingly important in practice as LLM-based coding tools are used by people with varying levels of programming expertise, including users who may only briefly inspect generated code~\cite{sarkarVibeCodingProgramming2025}.
Such errors may be difficult for LLMs to identify directly because they can depend on fine-grained token-level distinctions that are not always represented cleanly by model tokenisation~\cite{chaiTokenizationFallingShort2024}.
At the same time, LLMs are often effective at describing the higher-level intent of code in natural language~\cite{sunSourceCodeSummarization2025}.
This prompts a simple question: can code summaries help LLMs take a step back and recover behaviour-level context that is useful for repair?

Previous work has shown that natural-language summaries can act as useful intermediate artefacts for downstream software engineering tasks, including code rewriting and modification~\cite{liRewritingCodeSimple2024,tangExploringDirectInstruction2025}.
Building on this idea, we study summaries as lightweight diagnostic artefacts for repair workflows.
Rather than competing with full APR pipelines, we study a deliberately weak-specification setting: the model is given only the buggy code, not the original task description or a formal specification, and tests are used only as an evaluation oracle.
This can also reflect repair scenarios where users may know that the code is wrong, but may not be able to provide a precise technical diagnosis themselves~\cite{meskeVibeCodingReconfiguration2025}.
We test the diagnostic value of summaries through repair: if the summary helps recover the intended behaviour, then conditioning generation on it should make correct repairs more likely.

We call this pattern \textit{\textbf{summary-mediated repair}}.
The method is intentionally simple.
First, an LLM summarises the buggy code.
Then, the same LLM is asked to generate a repaired implementation conditioned on that summary.
By keeping the pipeline prompt-only, we avoid introducing additional tooling, retrieval, or agentic orchestration, and can more directly study whether natural-language summaries provide useful repair context.
The resulting summary also gives a human-readable view of what the model believes the code is doing, making the repair process more inspectable than directly generating a patch alone.

We evaluate summary-mediated repair across eight production-grade LLMs on two function-level repair settings.
First, we use existing bugs from \hepack~\cite{muennighoffOctoPackInstructionTuning2024}, allowing us to study whether summaries help repair externally provided buggy programs.
Second, we use each model's own failed generations on \mbpp~\cite{austinProgramSynthesisLarge2021}, allowing us to study whether summaries help with self-repair.
Across both settings, we compare several summary styles, including concise, intent-focused, and error-aware summaries, against a direct repair baseline.

Our results show that summaries can help, but only under the right conditions.
Error-aware diagnostic summaries perform best, repairing up to 65\% of \hepack\ bugs and improving over direct repair by 5\% on average.
Generic summaries are less reliable, and can even distract the model from the true fault.
Self-repair is substantially harder: both direct repair and summary-mediated repair struggle to repair models' own failed generations, with at most 11\% of errors repaired.
This suggests that summaries are most useful when they recover behaviour-level context for a newly seen bug, rather than when the original failure already reflects a deeper misunderstanding of the task.

Overall, our findings position code summaries as useful but limited diagnostic artefacts for LLM-based program repair.
They are simple to generate, easy for humans to inspect, and can improve repair when they encourage the model to reason explicitly about likely faults.
However, they are not a replacement for stronger repair signals such as tests, specifications, execution feedback, or more sophisticated repair pipelines.
Instead, our results suggest that summaries are best viewed as lightweight diagnostic context that can be integrated into broader APR workflows, and as a possible interaction pattern for natural-language-mediated coding tools.

\pagebreak

\noindent
\textit{Our contributions are as follows:}

\begin{enumerate}[left=0pt]

\item We study whether natural-language code summaries provide useful intermediate diagnostic context for LLM-based program repair, using \textit{summary-mediated repair} as a simple prompt-only test case.

\item We evaluate this pattern across eight LLMs and two function-level repair settings, covering both externally provided bugs and LLM self-repair.

\item We show that error-aware diagnostic summaries are more effective than generic summaries, producing modest but consistent gains over direct repair for external bugs, while self-repair remains challenging.

\item We release our code, prompts, full results, and evaluation harness publicly to support further work on summary-based diagnostic context for APR.\footnote{\repourl}

\end{enumerate}

\section{Related Work}\label{sec:related}

\paragraph{Code Generation}

LLMs trained on massive code corpora have been shown to excel at code generation~\cite{jiangSurveyLargeLanguage2024}, achieving impressive pass rates on functional benchmarks such as \humaneval and powering tools such as GitHub Copilot~\cite{chenEvaluatingLargeLanguage2021}.
However, further evaluations reveal correctness gaps: LLMs frequently fail on corner cases, produce subtle semantic errors, and replicate bugs from training data~\cite{chenDeepDiveLarge2025}.
Benchmarks that emphasise real-world scenarios -- such as \hepack~\cite{muennighoffOctoPackInstructionTuning2024} and \texttt{SWE-Bench}~\cite{jimenezSWEbenchCanLanguage2024a} -- highlight these limitations and motivate research into methods that improve the robustness of generated code.
These limitations are particularly important as LLM-based coding tools are increasingly used through natural-language interaction, including by users who may not have the expertise to inspect or debug generated code directly~\cite{sarkarVibeCodingProgramming2025}.

\paragraph{Code Summarisation}

Source code summarisation has long been studied as a way to automatically document and explain code~\cite{stapletonHumanStudyComprehension2020}.
Recent work shows that modern LLMs can produce accurate intent-capturing summaries of code~\cite{sunSourceCodeSummarization2025}.
LLM-produced summaries have already been used as intermediate artefacts for downstream tasks such as style normalisation~\cite{liRewritingCodeSimple2024} and developer-mediated code modification~\cite{tangExploringDirectInstruction2025}.
These findings motivate our investigation of whether code summaries can also serve as semantically rich intermediate artefacts for the task of program repair.

\paragraph{Code Repair}

Automated program repair (APR) has been extensively studied, evolving from search- and genetic-based methods to template- and constraint-based methods~\cite{huangSurveyAutomatedProgram2023,legouesAutomatedProgramRepair2019}, and more recently to LLM-driven approaches~\cite{yangSurveyLLMbasedAutomated2025}.
Despite these advances, persistent problems -- notably ``plausible-but-incorrect'' patches and weak fault localization -- continue to make small-scale semantic fixes unreliable~\cite{tanAntipatternsSearchbasedProgram2016}.
In addition, many techniques can struggle when there only exists a weak oracle to check for correctness~\cite{yeAutomatedPatchAssessment2021}.
Closest to our work, round-trip translation has been studied for APR, translating buggy code into another programming or natural language and back again to expose latent repair capabilities in LLMs~\cite{vallecillos-ruizAssessingLatentAutomated2026}.
Our work differs by isolating natural-language code summaries as the intermediate artefact, and by comparing how different summary styles affect repair.
This motivates exploration of lightweight repair techniques that can help identify subtle errors, provide interpretable diagnostic context, and remain useful when a formal specification of the intended behaviour is unavailable.

\section{Summary-Mediated Repair} \label{sec:method}

Our method is deliberately simple: we use code summarisation as an intermediate stage in a prompt-only APR process that does not require a formal specification.
A typical APR process takes the original $code$ as input, along with some $error$ description, produces a candidate $code'=\repair(code, error)$, and validates it on some test harness to get $result=\test(code')$.
Our \repair method has two stages and only requires the original $code$, as shown in Figure~\ref{fig:method}.
We first prompt an LLM to generate a summary, $summary=\summarise(code)$, describing the apparent behaviour of $code$.
We then prompt the LLM to generate a repaired program using the $summary$, $code'=\generate(summary)$, and validate on the same test harness to get $result=\test(code')$.

There is flexibility in the exact prompts used in both stages:
\summarise could request a concise one-line summary, a detailed list of behaviour, or an explicitly error-aware summary;
\generate could use a basic prompt, or provide more detailed instructions on how to interpret the summary.
In this paper, we focus on a deliberately weak-specification version of this pattern, where the pipeline receives only the buggy code and no task description, tests, traces, or formal specification.
This keeps the setting simple and allows us to study whether the summary itself provides useful repair context, rather than introducing additional signal through tooling, retrieval, execution feedback, or agentic orchestration.
It also reflects a practical use case for generative coding tools, where users increasingly write code through natural language and may not be able to provide a formal specification or precise diagnosis when it goes wrong~\cite{sarkarVibeCodingProgramming2025}.

The summary acts as a natural-language interaction point between the buggy code and the repair process, giving a readable account of what the LLM believes the code does before regeneration.
This is unlikely to fix all bug types, but may help when small implementation-level mistakes have large behavioural effects: the summary can encourage the model to recover the intended behaviour without directly copying the faulty surface form of the original code.
We therefore use this simple pipeline to test whether summaries provide useful behaviour-level diagnostic context when stronger repair signals are limited or unavailable.

\section{Experimental Design} \label{sec:experiment}

This section states the research questions (RQs) and describes the experiments we run to evaluate the effectiveness of \textit{summary-mediated repair} for repairing function-level errors.

\subsection{Research Questions} \label{sec:rqs}

\begin{rqlist}[left=5pt]

\item \textbf{Bug Repair}:
\textit{How effective is summary-mediated repair on existing buggy code when using different styles of summaries?}
To answer this RQ, we use \textit{summary-mediated repair} to fix buggy code from the \hepack dataset.

\item \textbf{Self Repair}: \textit{How effective is summary-mediated repair at helping an LLM to fix errors in its own code when using different styles of summaries?}
To answer this RQ, we ask different LLMs to solve tasks from \mbpp, and use \textit{summary-mediated repair} to try and fix any incorrect solutions.

\end{rqlist}

\subsection{LLM Selection}

We use a range of LLMs to assess \textit{summary-mediated repair} and enable a broad understanding of its effectiveness.
In particular, we want to experiment on a range of LLM sizes (number of parameters), different use-cases (general or code-specific), varying availabilities (open or closed source), and built by different organisations.
Therefore, we choose eight different models from four different providers (OpenAI, Llama, Mistral and Qwen). \textit{LLM details given in Table~\ref{tab:models}.}

\paragraph{LLM Configuration}

As is standard in code generation evaluation, we focus on low-variance single sample metrics (such as \textit{pass@1}~\cite{paulBenchmarksMetricsEvaluations2024b}) and therefore use conservative decoding settings (\(temperature=0.2\) and \(top\_p=1.0\)) to reduce sampling noise and make LLM comparisons fair.
A low, non-zero \(temperature\) value reduces randomness in responses and is a suitable setting when comparing code generation performance for single-sample runs~\cite{chenEvaluatingLargeLanguage2021};
and setting $top\_p=1.0$ disables nucleus truncation, ensuring decoding behaviour is primarily controlled by \(temperature\)~\cite{holtzmanCuriousCaseNeural2019}.
We conduct each LLM interaction in a fresh API session to avoid prompt caching between tasks~\cite{guAuditingPromptCaching2025};
and we do not use a system prompt to ensure that each LLM has its base functionality tested~\cite{muCloserLookSystem2025}.

\begin{table}[t]
    \caption{
        \textbf{\textit{Details of all LLMs used in this study.}}
        Entries marked with ``--'' indicate that the information was not available at the time of the study (June 2026).
    }
    \label{tab:models}
    
    \begin{adjustbox}{width=\columnwidth}
    \centering
    
\bgroup
\def\arraystretch{2.5}
    
    \begin{tabular}{ccccccc}
    \toprule
        \thead{Model} & \thead{Version} & \thead{Release} & \thead{Knowledge\\cut-off} & \thead{Size} & \thead{Open-\\source?} & \thead{Code\\model?} \\

    \midrule
        \makecell[c]{GPT-4o-\\mini~\cite{openaiGPT4oMiniAPI2025}} & \makecell[c]{gpt-4o-mini-\\2024-07-18} & July '24 & Oct. '23 & - & \xmark & \xmark \\
    
        \makecell[c]{GPT-4.1-\\mini~\cite{openaiGPT41MiniAPI2025}} & \makecell[c]{gpt-4.1-mini-\\2025-04-14} & Apr. '25 & Jun. '24 & - & \xmark & \xmark \\
    
        Llama-3.3~\cite{Llama33Model2025} & \makecell[c]{llama-3.3-70b-\\instruct-turbo} & Dec. '24 & Dec. '23 & 70B & \cmark & \xmark \\
    
        Llama-4~\cite{Llama4Model2025} & \makecell[c]{llama-4-scout-17b-\\16e-instruct} & Apr. '25 & Aug. '24 & 109B & \cmark & \xmark \\
    
        \makecell[c]{Qwen-2.5-\\Coder~\cite{huiQwen25CoderTechnicalReport2024}} & \makecell[c]{qwen2.5-coder-\\32b-instruct} & Sep. '24 & Mar. '24 & 32B & \cmark & \cmark \\
    
        \makecell[c]{Qwen-2.5-\\Turbo~\cite{qwenQwen25TechnicalReport2025}} & \makecell[c]{qwen2.5-72b-\\instruct-turbo} & Sep. '24 & Mar. '24 & 72B & \cmark & \xmark \\
    
        \makecell[c]{Codestral~\cite{Codestral2501Mistral2025}} & \makecell[c]{codestral-\\2501} & Jan. '25 & - & - & \xmark & \cmark \\
    
        \makecell[c]{Mistral-\\Medium~\cite{MediumNewLarge2025}} & \makecell[c]{mistral-\\medium-2505} & May '25 & - & - & \xmark & \xmark \\
    \bottomrule
    \end{tabular}

\egroup
    
    \end{adjustbox}
\end{table}

\subsection{Prompt Strategy}

Our \textit{summary-mediated repair} method (as defined in Section~\ref{sec:method}) intentionally separates the \summarise and \generate steps, and allows for a lot of flexibility in the prompts used in both.
We keep the \generate prompt consistent in all experiments and vary only the \summarise prompt.
This isolates the $summary$ artefact and reduces the number of variables, making it easier to attribute any observed differences to how the code is summarised, rather than the generation instructions.
We also explicitly choose function-level tasks for our experiments because they provide a controlled, reproducible setting that is appropriate for an initial, focused assessment of our method.
This setting does not capture the full complexity of repository-level APR, but allows us to isolate the contribution of the summary artefact itself.

\paragraph{Summarisation prompts}

We compare the effectiveness of five prompts for the \summarise step.
We do not aim to find the absolute best prompt; we sample sensible, interpretable prompt styles to get an initial assessment of how effective our method is and how impactful summary style is.
Together, these prompts allow us to compare neutral summaries, concise summaries, intent-focused summaries, and explicitly diagnostic summaries.
The summary prompts are as follows:

\noindent\textbf{Base --}
A neutral prompt with no extra guidance:
\prompt{Summarise the following python code: \textbf{\{code\}}}

\noindent\textbf{Short --}
Ask for a concise, one-sentence summary:
\prompt{Summarise the following python code in one sentence: \textbf{\{code\}}}

\noindent\textbf{Intent --}
Instruct the LLM to infer the original intent of the code:
\prompt{Summarise the following python code, be smart and infer expected functionality: \textbf{\{code\}}}

\noindent\textbf{Error --}
Explicitly tell the LLM that the code contains an error:
\prompt{Summarise the following python code, it contains at least one bug that is making a test fail: \textbf{\{code\}}}

\noindent\textbf{Warn --}
An error-aware prompt that warns about potential bugs:
\prompt{Summarise the following python code, it is not very well written, and might have some bugs, so please work around this when writing the summary: \textbf{\{code\}}}

\begin{table*}[ht]
    \caption{
        \textbf{\textit{Results for RQ1 Bug Repair.}}
        Results for LLMs fixing buggy code from the \hepack dataset using \textit{summary-mediated repair} with different styles of summary.
        For each LLM and repair method we give the \textit{fix@1} rate, the percentage of bugs fixed after one attempt of the repair method.
    }
    \label{tab:bug-repair}
    
    \centering
    \begin{adjustbox}{width=\textwidth}
    \begingroup
    \renewcommand{\arraystretch}{1.5} 

\begin{tabular}{m{1.5cm}m{1.0cm}m{2.0cm}m{2.0cm}m{2.0cm}m{2.0cm}m{2.0cm}m{2.0cm}m{2.0cm}m{2.0cm}}
\toprule
\multicolumn{2}{c}{\multirowcell{2}{\textbf{Repair Method}}} & \multicolumn{1}{c}{\textbf{GPT-4o-mini}} & \multicolumn{1}{c}{\textbf{GPT-4.1-mini}} & \multicolumn{1}{c}{\textbf{Llama-3.3-70B}} & \multicolumn{1}{c}{\textbf{Llama-4-Scout}} & \multicolumn{1}{c}{\textbf{Codestral}} & \multicolumn{1}{c}{\textbf{Mistral-Medium}} & \multicolumn{1}{c}{\textbf{Qwen-Coder}} & \multicolumn{1}{c}{\textbf{Qwen-Turbo}} \\
\multicolumn{2}{c}{} & \makecell[r]{{\textit{fix@1}}} & \makecell[r]{{\textit{fix@1}}} & \makecell[r]{{\textit{fix@1}}} & \makecell[r]{{\textit{fix@1}}} & \makecell[r]{{\textit{fix@1}}} & \makecell[r]{{\textit{fix@1}}} & \makecell[r]{{\textit{fix@1}}} & \makecell[r]{{\textit{fix@1}}} \\
\midrule
\multicolumn{2}{c}{Direct repair (baseline)} & \makecell[r]{54.88\%} & \makecell[r]{57.93\%} & \makecell[r]{42.07\%} & \makecell[r]{43.90\%} & \makecell[r]{51.83\%} & \makecell[r]{54.88\%} & \makecell[r]{52.44\%} & \makecell[r]{60.98\%} \\
\midrule
\multirowcell{5}{Summary-\\mediated\\repair} & Base & \makecell[r]{37.20\%} & \makecell[r]{26.22\%} & \makecell[r]{37.20\%} & \makecell[r]{39.02\%} & \makecell[r]{37.20\%} & \makecell[r]{37.20\%} & \makecell[r]{43.90\%} & \makecell[r]{45.12\%} \\

& Short & \makecell[r]{31.71\%} & \makecell[r]{22.56\%} & \makecell[r]{29.88\%} & \makecell[r]{39.02\%} & \makecell[r]{37.20\%} & \makecell[r]{35.37\%} & \makecell[r]{34.76\%} & \makecell[r]{40.85\%} \\

& Intent & \makecell[r]{33.54\%} & \makecell[r]{31.10\%} & \makecell[r]{37.20\%} & \makecell[r]{36.59\%} & \makecell[r]{39.63\%} & \makecell[r]{36.59\%} & \makecell[r]{47.56\%} & \makecell[r]{44.51\%} \\

& Error & \makecell[r]{57.32\%} & \makecell[r]{64.63\%} & \makecell[r]{52.44\%} & \makecell[r]{51.83\%} & \makecell[r]{57.32\%} & \makecell[r]{56.10\%} & \makecell[r]{58.54\%} & \makecell[r]{60.98\%} \\

& Warn & \makecell[r]{46.34\%} & \makecell[r]{33.54\%} & \makecell[r]{39.63\%} & \makecell[r]{37.20\%} & \makecell[r]{45.73\%} & \makecell[r]{41.46\%} & \makecell[r]{50.00\%} & \makecell[r]{56.71\%} \\
\hline
\end{tabular}

    \endgroup
    \end{adjustbox}
    
\end{table*}

\paragraph{Generation prompt}

After summarisation, we need to generate the updated code.
To ensure that the final output is directly executable by the unit-test harnesses used in our experiments, we use a simple \generate prompt that requests a function definition:

\prompt{Write a python function `\textbf{\{function\}}`, satisfying the following code summary: \textbf{\{summary\}}}

This intentionally makes the summary the primary repair context, allowing us to test whether the natural-language artefact alone captures enough behaviour-level information to support repair.

\paragraph{Baseline prompt}

We also need a baseline against which to compare our results.
Our method does not require the initial specification in order to attempt a repair; therefore, we do not include the initial specification for the baseline prompt either.
This gives the baseline and the full summary-mediated repair pipeline the same initial input: the buggy code only.
Our method first converts this code into a natural-language summary before generating a repair, while the baseline asks the LLM to repair the code directly.
We use a simple prompt that directly requests the LLM to repair a given piece of code:

\prompt{Repair the following python code: \textbf{\{code\}}}

\subsection{Experimental Setup for RQ1: Bug Repair}

For this RQ we want to investigate whether \textit{summary-mediated repair} is effective on existing code that contains bugs.
To do this we use the \hepack dataset~\cite{muennighoffOctoPackInstructionTuning2024} -- an extension of OpenAI's \humaneval~\cite{chenEvaluatingLargeLanguage2021} dataset that covers three distinct tasks, one of which is function-level APR.
\hepack provides a candidate solution (with a simple, manually inserted bug) and test cases for each of the 164 tasks from the original dataset.
For each task, we run the incorrect solution through our \textit{summary-mediated repair} pipeline (defined in Section~\ref{sec:method}), and validate the returned code using the provided tests.
The tests are used only for evaluation, and are not provided to the model during repair.
We consider a repair successful if all tests now pass.

We report single-sample results, defining the metric \textit{fix@1}: the percentage of bugs fixed after one execution of the \textit{summary-mediated repair} pipeline.

\subsection{Experimental Setup for RQ2: Self Repair}

For this RQ, we examine whether our approach can help LLMs to repair their own incorrect code generations.
We choose the \mbpp~\cite{austinProgramSynthesisLarge2021} dataset -- which contains 974 function-level \texttt{Python} programming tasks, each with a natural language task description and corresponding test cases -- and use the 500 tasks from its test split.
We first prompt the LLM to produce an initial solution and check correctness against the provided test cases.
If the solution fails, we apply the \textit{summary-mediated repair} pipeline (defined in Section~\ref{sec:method}), using the failing code as input to generate a new version.
As in RQ1, the original task description and tests are not provided during repair; they are used only to identify failing generations and evaluate the repaired code.
We again evaluate this new code by using the corresponding test cases to check for correctness.

We report two metrics: \textit{fix@1}, defined as the proportion of initially incorrect solutions that are successfully repaired after one execution of the \textit{summary-mediated repair} pipeline; and an adjusted \textit{pass@1}, representing the overall proportion of tasks correctly solved after the repair process has been applied.

\section{Results}\label{sec}

This section presents the results of our experiments into the effectiveness of \textit{summary-mediated repair}.

\subsection{Results for RQ1: Bug Repair}

This RQ investigates whether \textit{summary-mediated repair} is effective on existing buggy code.
Table~\ref{tab:bug-repair} reports the \textit{fix@1} rates for each LLM and repair method on the \hepack dataset.

\textit{Summary-mediated repair} shows potential, but its effectiveness depends strongly on the type of summary produced.
All summary prompts were able to repair at least 22.6\% of errors, but their effects are inconsistent.
The error-aware prompts (\textit{error} and \textit{warn}), where the LLM is told to expect errors in the code, provide the best results, with \textit{fix@1} rates of 57.4\% and 43.8\% respectively.
Yet, only the \textit{error} summary outperformed the \textit{direct repair} baseline, doing so by 5.0\% on average.
This indicates that summaries are most useful when they act as diagnostic artefacts, rather than neutral descriptions of the code; and that asking the LLM to treat the input as potentially incorrect produces more useful summaries for repair.

\begin{table*}[ht]
    \caption{
        \textbf{\textit{Results for RQ2 Self Repair.}}
        Results for LLMs solving tasks from the \mbpp dataset, then fixing any errors using \textit{summary-mediated repair} with different styles of summary.
        For each LLM and repair method we give the \textit{fix@1} rate, the percentage of initially incorrect solutions fixed with one attempt of the repair method; and an adjusted \textit{pass@1} rate, the overall proportion of tasks correctly solved after repair.
    }
    \label{tab:self-repair}
    
    \centering
    \begin{adjustbox}{width=\textwidth}
    \begingroup
    \renewcommand{\arraystretch}{1.5} 

\begin{tabular}{m{1.5cm}m{1.0cm}m{1.0cm}m{1.0cm}m{1.0cm}m{1.0cm}m{1.0cm}m{1.0cm}m{1.0cm}m{1.0cm}m{1.0cm}m{1.0cm}m{1.0cm}m{1.0cm}m{1.0cm}m{1.0cm}m{1.0cm}m{1.0cm}}
\toprule
\multicolumn{2}{c}{\multirowcell{2}{\textbf{Repair Method}}} & \multicolumn{2}{c}{\textbf{GPT-4o-mini}} & \multicolumn{2}{c}{\textbf{GPT-4.1-mini}} & \multicolumn{2}{c}{\textbf{Llama-3.3-70B}} & \multicolumn{2}{c}{\textbf{Llama-4-Scout}} & \multicolumn{2}{c}{\textbf{Codestral}} & \multicolumn{2}{c}{\textbf{Mistral-Medium}} & \multicolumn{2}{c}{\textbf{Qwen-Coder}} & \multicolumn{2}{c}{\textbf{Qwen-Turbo}} \\
\multicolumn{2}{c}{} & \makecell[r]{{\textit{fix@1}}} & \makecell[r]{{\textit{pass@1}}} & \makecell[r]{{\textit{fix@1}}} & \makecell[r]{{\textit{pass@1}}} & \makecell[r]{{\textit{fix@1}}} & \makecell[r]{{\textit{pass@1}}} & \makecell[r]{{\textit{fix@1}}} & \makecell[r]{{\textit{pass@1}}} & \makecell[r]{{\textit{fix@1}}} & \makecell[r]{{\textit{pass@1}}} & \makecell[r]{{\textit{fix@1}}} & \makecell[r]{{\textit{pass@1}}} & \makecell[r]{{\textit{fix@1}}} & \makecell[r]{{\textit{pass@1}}} & \makecell[r]{{\textit{fix@1}}} & \makecell[r]{{\textit{pass@1}}} \\
\midrule
\multicolumn{2}{c}{Solve rate (no repair)} & \makecell[r]{--} & \makecell[r]{66.60\%} & \makecell[r]{--} & \makecell[r]{75.20\%} & \makecell[r]{--} & \makecell[r]{67.00\%} & \makecell[r]{--} & \makecell[r]{65.40\%} & \makecell[r]{--} & \makecell[r]{66.60\%} & \makecell[r]{--} & \makecell[r]{70.80\%} & \makecell[r]{--} & \makecell[r]{77.80\%} & \makecell[r]{--} & \makecell[r]{75.20\%} \\
\multicolumn{2}{c}{Direct repair (baseline)} & \makecell[r]{3.59\%} & \makecell[r]{67.80\%} & \makecell[r]{3.23\%} & \makecell[r]{76.00\%} & \makecell[r]{4.85\%} & \makecell[r]{68.60\%} & \makecell[r]{2.31\%} & \makecell[r]{66.20\%} & \makecell[r]{2.99\%} & \makecell[r]{67.60\%} & \makecell[r]{4.79\%} & \makecell[r]{72.20\%} & \makecell[r]{10.81\%} & \makecell[r]{80.20\%} & \makecell[r]{4.03\%} & \makecell[r]{76.20\%} \\
\midrule
\multirowcell{5}{Summary-\\mediated\\repair} & Base & \makecell[r]{1.20\%} & \makecell[r]{67.00\%} & \makecell[r]{1.61\%} & \makecell[r]{75.60\%} & \makecell[r]{2.42\%} & \makecell[r]{67.80\%} & \makecell[r]{1.73\%} & \makecell[r]{66.00\%} & \makecell[r]{1.20\%} & \makecell[r]{67.00\%} & \makecell[r]{1.37\%} & \makecell[r]{71.20\%} & \makecell[r]{6.31\%} & \makecell[r]{79.20\%} & \makecell[r]{0.81\%} & \makecell[r]{75.40\%} \\

& Short & \makecell[r]{3.59\%} & \makecell[r]{67.80\%} & \makecell[r]{1.61\%} & \makecell[r]{75.60\%} & \makecell[r]{3.64\%} & \makecell[r]{68.20\%} & \makecell[r]{4.05\%} & \makecell[r]{66.80\%} & \makecell[r]{5.39\%} & \makecell[r]{68.40\%} & \makecell[r]{4.11\%} & \makecell[r]{72.00\%} & \makecell[r]{8.11\%} & \makecell[r]{79.60\%} & \makecell[r]{4.03\%} & \makecell[r]{76.20\%} \\

& Intent & \makecell[r]{1.80\%} & \makecell[r]{67.20\%} & \makecell[r]{1.61\%} & \makecell[r]{75.60\%} & \makecell[r]{1.82\%} & \makecell[r]{67.60\%} & \makecell[r]{2.31\%} & \makecell[r]{66.20\%} & \makecell[r]{1.20\%} & \makecell[r]{67.00\%} & \makecell[r]{2.74\%} & \makecell[r]{71.60\%} & \makecell[r]{3.60\%} & \makecell[r]{78.60\%} & \makecell[r]{0.00\%} & \makecell[r]{75.20\%} \\

& Error & \makecell[r]{4.79\%} & \makecell[r]{68.20\%} & \makecell[r]{4.84\%} & \makecell[r]{76.40\%} & \makecell[r]{3.64\%} & \makecell[r]{68.20\%} & \makecell[r]{3.47\%} & \makecell[r]{66.60\%} & \makecell[r]{1.80\%} & \makecell[r]{67.20\%} & \makecell[r]{5.48\%} & \makecell[r]{72.40\%} & \makecell[r]{9.91\%} & \makecell[r]{80.00\%} & \makecell[r]{8.06\%} & \makecell[r]{77.20\%} \\

& Warn & \makecell[r]{3.59\%} & \makecell[r]{67.80\%} & \makecell[r]{2.42\%} & \makecell[r]{75.80\%} & \makecell[r]{3.03\%} & \makecell[r]{68.00\%} & \makecell[r]{4.05\%} & \makecell[r]{66.80\%} & \makecell[r]{1.20\%} & \makecell[r]{67.00\%} & \makecell[r]{2.05\%} & \makecell[r]{71.40\%} & \makecell[r]{8.11\%} & \makecell[r]{79.60\%} & \makecell[r]{3.23\%} & \makecell[r]{76.00\%} \\
\hline
\end{tabular}

    \endgroup
    \end{adjustbox}
    
\end{table*}

The effectiveness of \textit{summary-mediated repair} is also LLM dependent and does not appear to be linked to the baseline \textit{fix@1} rate for \textit{direct repair}.
Qwen LLMs have the highest average \textit{fix@1} rates across summary styles (47.0\% for Qwen-Coder and 49.6\% for Qwen-Turbo), but they also have high baselines.
Llama LLMs on the other hand have some of the lowest average \textit{fix@1} rates (39.3\% for Llama-3.3-70B and 40.7\% for Llama-4-Scout), but they are the most comparable to their baselines.
In general, we observe no clear correlation between an LLM's ability to directly repair the code, and the effectiveness of our method.

\paragraph{Case analysis}

To understand when summary-mediated repair is most effective, we examine \hepack records that every summary variant successfully fixes but the \textit{direct repair} baseline fails to repair.
There are 24 such cases.
The most common underlying bug types are \textit{value misuse} (9 cases, typically involving subtle mistakes in how variables are interpreted) and \textit{excess logic} (6 cases, involving unnecessary or misleading logical steps).
These instances highlight scenarios where the summary appears to provide useful behaviour-level context, helping the model recover the intended functionality rather than overfitting to the buggy implementation.
The summaries steer the LLM toward the core behaviour, whereas the baseline prompt often misidentifies the bug, for example, focussing on output formatting rather than a subtle logical flaw.

\rqanswerbox{
\textit{\textbf{Findings for RQ1 Bug Repair.}}
\textit{Summary-mediated repair} is useful when the summary is diagnostic rather than
purely descriptive.
Error-aware summaries perform best, fixing 57.4\% of \hepack\ bugs on average across LLMs and improving over direct repair by 5.0 percentage points.
However, the gains are summary- and model-dependent, so summaries should be treated as lightweight diagnostic context rather than a standalone repair solution.
}

\subsection{Results for RQ2: Self Repair}

This RQ investigates whether \textit{summary-mediated repair} is effective when helping an LLM to fix errors in its own generated code.
Table~\ref{tab:self-repair} reports the \textit{fix@1} and adjusted \textit{pass@1} rates for each LLM and repair method on the \mbpp dataset.

Generally, LLMs struggled to repair their own code with both \textit{direct repair} prompts and our \textit{summary-mediated repair} method.
We observed much lower \textit{fix@1} rates for all summaries and LLMs (all less than 10\%, most less than 5\%).
The \textit{error} summary still performs best, fixing an average of 5.2\% of code errors across LLMs.
Perhaps surprisingly, the \textit{short} summary performs second best, suggesting that concise restatements may sometimes be more useful than detailed summaries when the model is simply taking another attempt at the task.
Overall, these improvements are small, and much more similar to the \textit{direct repair} baseline than for RQ1.

Again, the effects are highly dependent on the LLM and summary prompt in question.
Qwen-Coder has by far the highest success rate at fixing its own bugs, with an average \textit{fix@1} rate of 7.2\% across summaries -- more than double any other LLM.
Some LLM/prompt combinations do show meaningful improvements though: Codestral has a \textit{fix@1} rate of 5.4\% for the \textit{short} summary, up from a 3.0\% baseline and 1.4\% average across the other summaries; and Qwen-Turbo has a \textit{fix@1} rate of 8.1\% for the \textit{error} summary, when other summaries show little to no ability to fix bugs.

\paragraph{Case analysis}

For \mbpp, we found no cases in which a failed task was repaired by all summary variants but not by the \textit{direct repair} baseline, preventing a meaningful case analysis.
In general, few methods performed well on any given task, and there was no consistent pattern indicating when \textit{summary-mediated repair} outperformed the baseline.
In most cases, the LLM either recognised its own error or it did not, and the specific prompting strategy had minimal influence on this outcome.
This supports the view that summaries are most useful when they add behaviour-level diagnostic context; for self-repair, the summary is often derived from the same misunderstanding that caused the original failure.

\rqanswerbox{
\textit{\textbf{Findings for RQ2 Self Repair.}}
\textit{Summary-mediated repair} fixes only a small fraction of initially failed \mbpp\ generations, with the best summary style repairing 5.2\% of failures on average.
This suggests that summaries help most when they recover useful behaviour-level context, but provide little new signal when the original code already reflects a misunderstanding of the task.
}

\section{Discussion}\label{sec:future}




\paragraph{Practical Implications}

Our experiments show that natural language code summaries have potential as lightweight intermediate artefacts in APR pipelines.
They do not achieve state-of-the-art APR on their own, and should not be seen as a replacement for stronger repair signals such as tests, traces, or fault localisation.
Instead, they provide a simple way to add diagnostic context to LLM-based repair, particularly when a formal specification or precise error description is unavailable.
Error-aware summaries--those that ask the LLM to expect errors or flag suspicious code--produce the best results;
naive summaries, by contrast, can strip away useful implementation detail and hurt repair.
This suggests that summaries are most useful when they help the model reason about likely faults, rather than merely restating what the code appears to do.

The self-repair results are more limited.
When an LLM fails to solve a task initially, \textit{summary-mediated repair} usually yields only marginal gains.
This suggests that the method can help expose an LLM's existing ability to repair newly seen bugs, but is unlikely to substantially improve its general ability to solve coding problems.
Practically, summaries are simple to generate, easy for humans to inspect, and most useful alongside complementary techniques such as behavioural checks, execution feedback, or human-in-the-loop adjustment.
They may also provide a useful natural-language interaction point in coding tools, helping users understand what the model believes the code is doing before accepting a patch.

\paragraph{Limitations}

There are several important limitations to be aware of.
First, we intentionally focus on function-level tasks because these problems are small and self-contained, allowing controlled evaluation;
the results may not generalise to project-level bugs.
Second, our repair setting is deliberately weak-specification: the model receives the buggy code, but not the original task description, failing tests, execution traces, or a formal specification.
This isolates the contribution of the summary artefact, but also limits what the model can infer.
Third, our metrics are single-sample (\textit{fix@1}, \textit{pass@1}) and therefore sensitive to decoding and sampling choices;
we mitigate this with conservative decoding settings, but more extensive experiments, including repeated runs, confidence intervals, significance tests, and \textit{pass@k}, could change effect sizes.
Finally, passing unit tests does not guarantee semantic correctness: plausible-but-incorrect patches remain a concern.
Larger manual audits would help clarify exactly where this process assists LLMs.

\paragraph{Future work}

Future work should explore improved summary designs and broader evaluation settings.
Our experiments suggest that error-aware and concise summaries can help in different ways;
combining these styles, for example through short diagnostic summaries, may provide a useful balance between focus and fault awareness.
It would also be valuable to study summary-mediated repair with stronger repair signals, such as failing test outputs, execution traces, or lightweight fault localisation.
Finally, the method should be evaluated in more realistic settings, including industrial function-level bugs and project-level errors.
More broadly, our results add to evidence that natural-language summaries can be useful intermediate artefacts for downstream coding tasks beyond documentation alone.

\section{Conclusion}\label{sec:conclusion}

We introduce \textit{summary-mediated repair}, a simple prompt-only repair pattern that asks an LLM to summarise buggy code and then regenerate the code conditioned on that summary.
We evaluate the technique across eight LLMs on function-level tasks.
\textit{Summary-mediated repair} shows potential as a lightweight diagnostic layer for LLM-based APR, but gains are modest and LLM-dependent;
only error-aware summaries consistently outperform direct repair.

Although not state-of-the-art on its own, \textit{summary-mediated repair} highlights the value of code summaries as interpretable intermediate artefacts for downstream coding tasks.
Summaries can sometimes provide useful behaviour-level context, make repair more inspectable, and offer a natural-language interaction point for coding tools.
We release the full code and results on our GitHub repository,\footnote{\repourl} and hope this work encourages further investigation into code summarisation as diagnostic context for program repair and other natural-language-mediated coding workflows.

\balance

\bibliographystyle{ACM-Reference-Format}
\bibliography{main}

@misc{openaiGPT4oMiniAPI2025,
  title = {{{GPT-4o}} Mini - {{API}}},
  author = {OpenAI},
  year = {2025},
  journal = {OpenAI Platform},
  howpublished = {https://platform.openai.com/docs/models/gpt-4o-mini},
  langid = {american}
}

@misc{openaiGPT41MiniAPI2025,
  title = {{{GPT-4}}.1 Mini - {{API}}},
  author = {OpenAI},
  year = {2025},
  journal = {OpenAI Platform},
  howpublished = {https://platform.openai.com/docs/models/gpt-4.1-mini},
  langid = {american}
}

@misc{Codestral2501Mistral2025,
  title = {Codestral 25.01 {\textbar} {{Mistral AI}}},
  author = {{Mistral AI}},
  year = {2025},
  journal = {Mistral AI},
  howpublished = {https://mistral.ai/news/codestral-2501},
  langid = {english}
}

@misc{MediumNewLarge2025,
  title = {Medium Is the New Large. {\textbar} {{Mistral AI}}},
  author = {{Mistral AI}},
  year = {2025},
  journal = {Mistral AI},
  howpublished = {https://mistral.ai/ news/mistral-medium-3},
  langid = {english}
}

@misc{qwenQwen25TechnicalReport2025,
  title = {Qwen2.5 {{Technical Report}}},
  author = {Qwen and Yang, An and Yang, Baosong and Zhang, Beichen and Hui, Binyuan and Zheng, Bo and Yu, Bowen and Li, Chengyuan and Liu, Dayiheng and Huang, Fei and Wei, Haoran and Lin, Huan and Yang, Jian and Tu, Jianhong and Zhang, Jianwei and Yang, Jianxin and Yang, Jiaxi and Zhou, Jingren and Lin, Junyang and Dang, Kai and Lu, Keming and Bao, Keqin and Yang, Kexin and Yu, Le and Li, Mei and Xue, Mingfeng and Zhang, Pei and Zhu, Qin and Men, Rui and Lin, Runji and Li, Tianhao and Tang, Tianyi and Xia, Tingyu and Ren, Xingzhang and Ren, Xuancheng and Fan, Yang and Su, Yang and Zhang, Yichang and Wan, Yu and Liu, Yuqiong and Cui, Zeyu and Zhang, Zhenru and Qiu, Zihan},
  year = {2025},
  month = jan,
  number = {arXiv:2412.15115},
  eprint = {2412.15115},
  primaryclass = {cs},
  publisher = {arXiv},
  doi = {10.48550/arXiv.2412.15115},
  archiveprefix = {arXiv}
}

@misc{huiQwen25CoderTechnicalReport2024,
  title = {Qwen2.5-{{Coder Technical Report}}},
  author = {Hui, Binyuan and Yang, Jian and Cui, Zeyu and Yang, Jiaxi and Liu, Dayiheng and Zhang, Lei and Liu, Tianyu and Zhang, Jiajun and Yu, Bowen and Lu, Keming and Dang, Kai and Fan, Yang and Zhang, Yichang and Yang, An and Men, Rui and Huang, Fei and Zheng, Bo and Miao, Yibo and Quan, Shanghaoran and Feng, Yunlong and Ren, Xingzhang and Ren, Xuancheng and Zhou, Jingren and Lin, Junyang},
  year = {2024},
  month = nov,
  number = {arXiv:2409.12186},
  eprint = {2409.12186},
  primaryclass = {cs},
  publisher = {arXiv},
  doi = {10.48550/arXiv.2409.12186},
  archiveprefix = {arXiv}
}

@inproceedings{stapletonHumanStudyComprehension2020,
  title = {A {{Human Study}} of {{Comprehension}} and {{Code Summarization}}},
  booktitle = {Proceedings of the 28th {{International Conference}} on {{Program Comprehension}}},
  author = {Stapleton, Sean and Gambhir, Yashmeet and LeClair, Alexander and Eberhart, Zachary and Weimer, Westley and Leach, Kevin and Huang, Yu},
  year = {2020},
  month = sep,
  series = {{{ICPC}} '20},
  pages = {2--13},
  publisher = {Association for Computing Machinery},
  address = {New York, NY, USA},
  doi = {10.1145/3387904.3389258},
  isbn = {978-1-4503-7958-8}
}

@incollection{sunSourceCodeSummarization2025,
  title = {Source {{Code Summarization}} in the {{Era}} of {{Large Language Models}}},
  booktitle = {Proceedings of the {{IEEE}}/{{ACM}} 47th {{International Conference}} on {{Software Engineering}}},
  author = {Sun, Weisong and Miao, Yun and Li, Yuekang and Zhang, Hongyu and Fang, Chunrong and Liu, Yi and Deng, Gelei and Liu, Yang and Chen, Zhenyu},
  year = {2025},
  month = sep,
  pages = {1882--1894},
  publisher = {IEEE Press},
  isbn = {979-8-3315-0569-1}
}

@misc{tangExploringDirectInstruction2025,
  title = {Exploring {{Direct Instruction}} and {{Summary-Mediated Prompting}} in {{LLM-Assisted Code Modification}}},
  author = {Tang, Ningzhi and Smith, Emory and Huang, Yu and McMillan, Collin and Li, Toby Jia-Jun},
  year = {2025},
  month = aug,
  number = {arXiv:2508.01523},
  eprint = {2508.01523},
  primaryclass = {cs},
  publisher = {arXiv},
  doi = {10.48550/arXiv.2508.01523},
  archiveprefix = {arXiv}
}

@inproceedings{liRewritingCodeSimple2024,
  title = {Rewriting the {{Code}}: {{A Simple Method}} for {{Large Language Model Augmented Code Search}}},
  shorttitle = {Rewriting the {{Code}}},
  booktitle = {Proceedings of the 62nd {{Annual Meeting}} of the {{Association}} for {{Computational Linguistics}} ({{Volume}} 1: {{Long Papers}})},
  author = {Li, Haochen and Zhou, Xin and Shen, Zhiqi},
  editor = {Ku, Lun-Wei and Martins, Andre and Srikumar, Vivek},
  year = {2024},
  month = aug,
  pages = {1371--1389},
  publisher = {Association for Computational Linguistics},
  address = {Bangkok, Thailand},
  doi = {10.18653/v1/2024.acl-long.75},
}

@misc{chenDeepDiveLarge2025,
  title = {A {{Deep Dive Into Large Language Model Code Generation Mistakes}}: {{What}} and {{Why}}?},
  shorttitle = {A {{Deep Dive Into Large Language Model Code Generation Mistakes}}},
  author = {Chen, QiHong and Yu, Jiachen and Li, Jiawei and Deng, Jiecheng and Chen, Justin Tian Jin and Ahmed, Iftekhar},
  year = {2025},
  month = mar,
  number = {arXiv:2411.01414},
  eprint = {2411.01414},
  primaryclass = {cs},
  publisher = {arXiv},
  doi = {10.48550/arXiv.2411.01414},
  archiveprefix = {arXiv}
}

@article{jiangSurveyLargeLanguage2024,
  title = {A {{Survey}} on {{Large Language Models}} for {{Code Generation}}},
  author = {Jiang, Juyong and Wang, Fan and Shen, Jiasi and Kim, Sungju and Kim, Sunghun},
  year = {2024},
  month = jun,
  journal = {ACM Transactions on Software Engineering and Methodology},
  eprint = {2406.00515},
  archiveprefix = {arXiv}
}

@misc{chenEvaluatingLargeLanguage2021,
  title = {Evaluating {{Large Language Models Trained}} on {{Code}}},
  author = {Chen, Mark and Tworek, Jerry and Jun, Heewoo and Yuan, Qiming and Pinto, Henrique Ponde de Oliveira and Kaplan, Jared and Edwards, Harri and Burda, Yuri and Joseph, Nicholas and Brockman, Greg and Ray, Alex and Puri, Raul and Krueger, Gretchen and Petrov, Michael and Khlaaf, Heidy and Sastry, Girish and Mishkin, Pamela and Chan, Brooke and Gray, Scott and Ryder, Nick and Pavlov, Mikhail and Power, Alethea and Kaiser, Lukasz and Bavarian, Mohammad and Winter, Clemens and Tillet, Philippe and Such, Felipe Petroski and Cummings, Dave and Plappert, Matthias and Chantzis, Fotios and Barnes, Elizabeth and {Herbert-Voss}, Ariel and Guss, William Hebgen and Nichol, Alex and Paino, Alex and Tezak, Nikolas and Tang, Jie and Babuschkin, Igor and Balaji, Suchir and Jain, Shantanu and Saunders, William and Hesse, Christopher and Carr, Andrew N. and Leike, Jan and Achiam, Josh and Misra, Vedant and Morikawa, Evan and Radford, Alec and Knight, Matthew and Brundage, Miles and Murati, Mira and Mayer, Katie and Welinder, Peter and McGrew, Bob and Amodei, Dario and McCandlish, Sam and Sutskever, Ilya and Zaremba, Wojciech},
  year = {2021},
  month = jul,
  number = {arXiv:2107.03374},
  eprint = {2107.03374},
  publisher = {arXiv},
  doi = {10.48550/arXiv.2107.03374},
  archiveprefix = {arXiv}
}

@inproceedings{muennighoffOctoPackInstructionTuning2024,
  title = {{{OctoPack}}: {{Instruction Tuning Code Large Language Models}}},
  shorttitle = {{{OctoPack}}},
  booktitle = {The {{Twelfth International Conference}} on {{Learning Representations}}},
  author = {Muennighoff, Niklas and Liu, Qian and Zebaze, Armel and Zheng, Qinkai and Hui, Binyuan and Zhuo, Terry Yue and Singh, Swayam and Tang, Xiangru and von Werra, Leandro and Longpre, Shayne},
  year = {2024},
  month = feb,
  eprint = {2308.07124},
  primaryclass = {cs},
  doi = {10.48550/arXiv.2308.07124},
  archiveprefix = {arXiv}
}

@inproceedings{jimenezSWEbenchCanLanguage2024a,
  title = {{{SWE-bench}}: {{Can Language Models Resolve Real-World GitHub Issues}}?},
  shorttitle = {{{SWE-bench}}},
  booktitle = {The {{Twelfth International Conference}} on {{Learning Representations}}},
  author = {Jimenez, Carlos E. and Yang, John and Wettig, Alexander and Yao, Shunyu and Pei, Kexin and Press, Ofir and Narasimhan, Karthik},
  year = {2024},
  eprint = {2310.06770},
  primaryclass = {cs},
  publisher = {arXiv},
  doi = {10.48550/arXiv.2310.06770},
  archiveprefix = {arXiv}
}

@misc{huangSurveyAutomatedProgram2023,
  title = {A {{Survey}} on {{Automated Program Repair Techniques}}},
  author = {Huang, Kai and Xu, Zhengzi and Yang, Su and Sun, Hongyu and Li, Xuejun and Yan, Zheng and Zhang, Yuqing},
  year = {2023},
  month = may,
  number = {arXiv:2303.18184},
  eprint = {2303.18184},
  primaryclass = {cs},
  publisher = {arXiv},
  doi = {10.48550/arXiv.2303.18184},
  archiveprefix = {arXiv}
}

@misc{yangSurveyLLMbasedAutomated2025,
  title = {A {{Survey}} of {{LLM-based Automated Program Repair}}: {{Taxonomies}}, {{Design Paradigms}}, and {{Applications}}},
  shorttitle = {A {{Survey}} of {{LLM-based Automated Program Repair}}},
  author = {Yang, Boyang and Cai, Zijian and Liu, Fengling and Le, Bach and Zhang, Lingming and Bissyand{\'e}, Tegawend{\'e} F. and Liu, Yang and Tian, Haoye},
  year = {2025},
  month = jun,
  number = {arXiv:2506.23749},
  eprint = {2506.23749},
  primaryclass = {cs},
  publisher = {arXiv},
  doi = {10.48550/arXiv.2506.23749},
  archiveprefix = {arXiv}
}

@inproceedings{tanAntipatternsSearchbasedProgram2016,
  title = {Anti-Patterns in Search-Based Program Repair},
  booktitle = {Proceedings of the 2016 24th {{ACM SIGSOFT International Symposium}} on {{Foundations}} of {{Software Engineering}}},
  author = {Tan, Shin Hwei and Yoshida, Hiroaki and Prasad, Mukul R. and Roychoudhury, Abhik},
  year = {2016},
  month = nov,
  series = {{{FSE}} 2016},
  pages = {727--738},
  publisher = {Association for Computing Machinery},
  address = {New York, NY, USA},
  doi = {10.1145/2950290.2950295},
  isbn = {978-1-4503-4218-6}
}

@article{legouesAutomatedProgramRepair2019,
  title = {Automated Program Repair},
  author = {Le Goues, Claire and Pradel, Michael and Roychoudhury, Abhik},
  year = {2019},
  month = nov,
  journal = {Commun. ACM},
  volume = {62},
  number = {12},
  pages = {56--65},
  issn = {0001-0782},
  doi = {10.1145/3318162},
}

@inproceedings{chaiTokenizationFallingShort2024,
  title = {Tokenization {{Falling Short}}: {{On Subword Robustness}} in {{Large Language Models}}},
  shorttitle = {Tokenization {{Falling Short}}},
  booktitle = {Findings of the {{Association}} for {{Computational Linguistics}}: {{EMNLP}} 2024},
  author = {Chai, Yekun and Fang, Yewei and Peng, Qiwei and Li, Xuhong},
  editor = {{Al-Onaizan}, Yaser and Bansal, Mohit and Chen, Yun-Nung},
  year = {2024},
  month = nov,
  pages = {1582--1599},
  publisher = {Association for Computational Linguistics},
  address = {Miami, Florida, USA},
  doi = {10.18653/v1/2024.findings-emnlp.86},
}

@inproceedings{guAuditingPromptCaching2025,
  title = {Auditing {{Prompt Caching}} in {{Language Model APIs}}},
  booktitle = {Forty-Second {{International Conference}} on {{Machine Learning}}},
  author = {Gu, Chenchen and Li, Xiang Lisa and Kuditipudi, Rohith and Liang, Percy and Hashimoto, Tatsunori},
  year = {2025},
  month = feb,
  eprint = {2502.07776},
  primaryclass = {cs},
  archiveprefix = {arXiv}
}

@inproceedings{muCloserLookSystem2025,
  title = {A {{Closer Look}} at {{System Prompt Robustness}}},
  booktitle = {Neurips {{Safe Generative AI Workshop}} 2024},
  author = {Mu, Norman and Lu, Jonathan and Lavery, Michael and Wagner, David},
  year = {2025},
  month = feb,
  eprint = {2502.12197},
  primaryclass = {cs},
  archiveprefix = {arXiv}
}

@inproceedings{holtzmanCuriousCaseNeural2019,
  title = {The {{Curious Case}} of {{Neural Text Degeneration}}},
  booktitle = {International {{Conference}} on {{Learning Representations}}},
  author = {Holtzman, Ari and Buys, Jan and Du, Li and Forbes, Maxwell and Choi, Yejin},
  year = {2019},
  month = sep,
  langid = {english}
}

@misc{Llama33Model2025,
  title = {Llama 3.3 {\textbar} {{Model Cards}} and {{Prompt}} Formats},
  author = {Meta},
  year = {2025},
  journal = {Llama},
  howpublished = {https://www.llama.com/ docs/model-cards-and-prompt-formats/llama3\_3/},
  langid = {english}
}

@misc{Llama4Model2025,
  title = {Llama 4 {\textbar} {{Model Cards}} and {{Prompt}} Formats},
  author = {Meta},
  year = {2025},
  journal = {Llama},
  howpublished = {https://www.llama.com/ docs/model-cards-and-prompt-formats/llama4/},
  langid = {english}
}

@misc{austinProgramSynthesisLarge2021,
  title = {Program {{Synthesis}} with {{Large Language Models}}},
  author = {Austin, Jacob and Odena, Augustus and Nye, Maxwell and Bosma, Maarten and Michalewski, Henryk and Dohan, David and Jiang, Ellen and Cai, Carrie and Terry, Michael and Le, Quoc and Sutton, Charles},
  year = {2021},
  month = aug,
  eprint = {2108.07732},
  publisher = {arXiv},
  doi = {10.48550/arXiv.2108.07732},
  archiveprefix = {arXiv}
}

@inproceedings{paulBenchmarksMetricsEvaluations2024b,
  title = {Benchmarks and {{Metrics}} for {{Evaluations}} of {{Code Generation}}: {{A Critical Review}}},
  shorttitle = {Benchmarks and {{Metrics}} for {{Evaluations}} of {{Code Generation}}},
  booktitle = {2024 {{IEEE International Conference}} on {{Artificial Intelligence Testing}} ({{AITest}})},
  author = {Paul, Debalina Ghosh and Zhu, Hong and Bayley, Ian},
  year = 2024,
  month = jul,
  pages = {87--94},
  publisher = {IEEE Computer Society},
  doi = {10.1109/AITest62860.2024.00019},
  isbn = {979-8-3503-6505-4},
  langid = {english}
}

@article{chenSurveyEvaluatingLarge2024,
  title = {A {{Survey}} on {{Evaluating Large Language Models}} in {{Code Generation Tasks}}},
  author = {Chen, Liguo and Guo, Qi and Jia, Hongrui and Zeng, Zhengran and Wang, Xin and Xu, Yijiang and Wu, Jian and Wang, Yidong and Gao, Qing and Wang, Jindong and Ye, Wei and Zhang, Shikun},
  year = 2024,
  journal = {Journal of computer science and technology},
  publisher = {arXiv},
  doi = {10.48550/ARXIV.2408.16498},
  copyright = {Creative Commons Attribution 4.0 International}
}

@inproceedings{liuYourCodeGenerated2023,
  title = {Is Your Code Generated by {{ChatGPT}} Really Correct? Rigorous Evaluation of Large Language Models for Code Generation},
  shorttitle = {Is Your Code Generated by {{ChatGPT}} Really Correct?},
  booktitle = {Proceedings of the 37th {{International Conference}} on {{Neural Information Processing Systems}}},
  author = {Liu, Jiawei and Xia, Chunqiu Steven and Wang, Yuyao and Zhang, Lingming},
  year = 2023,
  month = dec,
  series = {{{NIPS}} '23},
  pages = {21558--21572},
  publisher = {Curran Associates Inc.},
  address = {Red Hook, NY, USA},
}

@article{yeAutomatedPatchAssessment2021,
  title = {Automated Patch Assessment for Program Repair at Scale},
  author = {Ye, He and Martinez, Matias and Monperrus, Martin},
  year = 2021,
  month = feb,
  journal = {Empirical Software Engineering},
  volume = {26},
  number = {2},
  pages = {20},
  issn = {1573-7616},
  doi = {10.1007/s10664-020-09920-w},
  langid = {english}
}

@article{meskeVibeCodingReconfiguration2025,
  title = {Vibe {{Coding}} as a {{Reconfiguration}} of {{Intent Mediation}} in {{Software Development}}: {{Definition}}, {{Implications}}, and {{Research Agenda}}},
  shorttitle = {Vibe {{Coding}} as a {{Reconfiguration}} of {{Intent Mediation}} in {{Software Development}}},
  author = {Meske, Christian and Hermanns, Tobias and {Von der Weiden}, Esther and Loser, Kai-Uwe and Berger, Thorsten},
  year = 2025,
  journal = {IEEE Access},
  volume = {13},
  pages = {213242--213259},
  issn = {2169-3536},
  doi = {10.1109/ACCESS.2025.3645466}
}

@inproceedings{sarkarVibeCodingProgramming2025,
  title = {Vibe Coding: Programming through Conversation with Artificial Intelligence},
  shorttitle = {Vibe Coding},
  booktitle = {Proceedings of the 36th {{Annual Conference}} of the {{Psychology}} of {{Programming Interest Group}}},
  author = {Sarkar, Advait and Drosos, Ian},
  year = 2025,
  month = jun,
  eprint = {2506.23253},
  primaryclass = {cs},
  archiveprefix = {arXiv}
}

@article{vallecillos-ruizAssessingLatentAutomated2026,
  title = {Assessing the {{Latent Automated Program Repair Capabilities}} of {{Large Language Models Using Round-Trip Translation}}},
  author = {{Vallecillos-Ruiz}, Fernando and Grishina, Anastasiia and Hort, Max and Moonen, Leon},
  year = 2026,
  month = jul,
  journal = {ACM Transactions on Software Engineering and Methodology},
  volume = {35},
  number = {7},
  pages = {203:1--203:37},
  issn = {1049-331X},
  doi = {10.1145/3771922}
}

\end{document}